\documentclass[journal]{IEEEtran}
\ifCLASSINFOpdf
\else
\fi
\usepackage{cite}
\usepackage[draft]{hyperref}
\usepackage{amsmath}
\usepackage{amsfonts}
\usepackage{graphicx}
\usepackage{multirow}
\usepackage{float}
\usepackage{wrapfig}
\usepackage{hyperref}
\usepackage{graphicx}
\usepackage{caption}
\usepackage{subcaption}
\usepackage{epstopdf}
\usepackage{color}
\usepackage{longtable}
\usepackage{moreverb}
\usepackage{amsmath}
\usepackage{algorithmicx}
\usepackage{algorithm}
\usepackage{algpseudocode}
\usepackage[labelsep=period]{caption}
\usepackage{textcase}

\algdef{SE}[DOWHILE]{Do}{doWhile}{\algorithmicdo}[1]{\algorithmicwhile\ #1}%
\usepackage{listings}
\usepackage{esvect}

\IEEEoverridecommandlockouts \IEEEpubid{\makebox[\columnwidth]{ 978-1-5386-3531-5/17/\$31.00~\copyright~2017 IEEE \hfill} \hspace{\columnsep}\makebox[\columnwidth]{ }}

\begin{document}
%
\title{UAV Assisted Public Safety Communications with LTE-Advanced HetNets and FeICIC}
%
%
%

\author{\IEEEauthorblockN{Abhaykumar Kumbhar$^{1,2}$, Simran Singh$^{3}$, and \.{I}smail~G\"uven\c{c}$^{3}$}\\
\IEEEauthorblockA{$^1$Dept. Electrical and Computer Engineering, Florida International University, Miami, FL, 33174\\
$^2$Motorola Solutions, Inc., Plantation, FL, 33322\\
$^3$Dept. Electrical and Computer Engineering, North Carolina State University, Raleigh, NC, 27606
\vspace{-0.5cm}
}}

\maketitle

\begin{abstract}
Establishing a reliable communication infrastructure at an emergency site is a crucial task for mission-critical and real-time public safety communications (PSC). To this end, use of unmanned aerial vehicles (UAVs) has recently received extensive interest for PSC to establish reliable connectivity in a heterogeneous network (HetNet) environment. These UAVs can be deployed as unmanned aerial base stations (UABSs) as part of the HetNet infrastructure. In this article, we explore the role of agile UABSs in LTE-Advanced HetNets by applying 3GPP Release-11 further-enhanced inter-cell interference coordination (FeICIC) and cell range expansion (CRE) techniques. Through simulations, we compare the system-wide 5th percentile spectral efficiency (SE) when UABSs are deployed in a hexagonal grid and when their locations are optimized using a genetic algorithm, while also jointly optimizing the CRE and the FeICIC parameters. Our simulation results show that at optimized UABS locations, the 3GPP Release-11 FeICIC with reduced power subframes can provide considerably better 5th percentile SE than the 3GPP Release-10 with almost blank subframes.
\end{abstract}


%
\IEEEpeerreviewmaketitle

\section{Introduction}
Public safety communications (PSC) is considered to be the cornerstone of public safety response system and plays a critical role in saving lives, property, and national infrastructure during a natural or man-made emergency. The legacy PSC technologies are designed predominantly for delivering mission-critical voice communications over narrowband channels, which have been so far met by operating in the pre-defined channelized spectrum allocation. However, the evolution of data and video applications demands higher channel capacity and improved spectral efficiency (SE)~\cite{R1,athukoralage2016regret}.

To enhance the capabilities of next-gen broadband PSC networks, recently, FirstNet in the United States is building a 4G Long Term Evolution (LTE) based coast-to-coast public safety network deployed in the 700~MHz band~\cite{R1}. Similarly, the United Kingdom plans to replace the TETRA system, which currently provides mission-critical communications for public safety agencies and other government organizations, with LTE by the year 2020~\cite{UKESNCurrentStatus}. The 4G mobile networks as considered in these examples have great potential to revolutionize PSC during emergency situations by providing high-speed real-time video and multimedia services along with mission-critical communication. Furthermore, LTE-Advanced capabilities such as small cell deployment, interference coordination, and cell range extension can restore or extend coverage beyond the existing or damaged PSC networks.

\begin{figure} [t]
\centering
\includegraphics[width=0.8\linewidth]{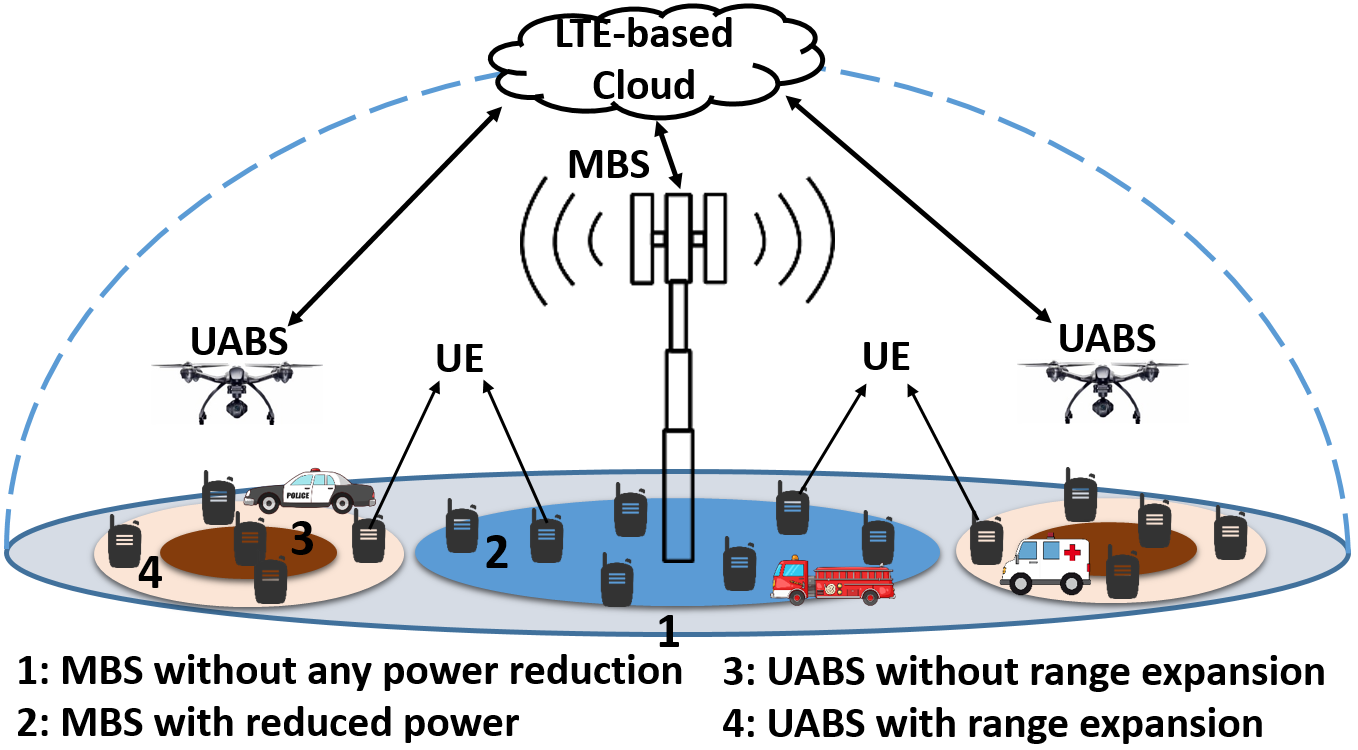}
\caption{The PSC scenario with MBS and UABSs constituting an air/ground HetNet infrastructure. The MBS can use the ICIC techniques defined in LTE-Advanced. The UABSs can dynamically change their position to maintain good coverage and can utilize range expansion bias to take over MBS UEs.}
\label{PscHetnet}\vspace{-0.45cm}
\end{figure}

Unmanned aerial base stations (UABSs) such as balloons, quadcopters, and gliders equipped with LTE-Advanced capabilities can be utilized for emergency restoration and temporary expansion of public safety network in case of disaster recovery~\cite{R4}. These UABSs can be deployed with minimum interdependencies, at low cost, and provide virtually omnipresent coverage which is essential for first-responders to be efficient and save lives. The FirstNet PSC network requires $95\%$ geographical coverage of the country, which will be difficult to achieve using only dedicated cell towers. The UABSs can be deployed when necessary to assist in achieving this coverage goal. On the other hand, the UABSs may also introduce significant interference problems with the ground network~\cite{moore2014first, merwaday2016improved,R10}.

Recent studies in the literature~\cite{gomez2016aerial,gomez2013performance, wang2016hybrid} have addressed the application of UABSs for rendering mission-critical communication. However, in a heterogeneous network (HetNet) environment,  use of UABSs introduce only limited performance gains due to high inter-cell interference. Deployment of unmanned aerial vehicles (UAVs) as mobile LTE relays to offload traffic in a HetNet scenario while also considering inter-cell interference has been studied in~\cite{rohde2013ad,sharma2017intelligent}. The effectiveness of 3GPP Release-10/11 inter-cell interference coordination (ICIC) techniques for fixed HetNet deployments has been explored in~\cite{R10}, while the use of cell range expansion (CRE) techniques for offloading users from MBSs to UABSs has been analyzed in~\cite{merwaday2016improved} without considering ICIC. To our best knowledge, merits of 3GPP Release-10/11 techniques along with CRE and UABS mobility have not been evaluated in the literature, and such an evaluation is the main goal of this paper. We consider an LTE band class~14 PSC network~\cite{R1} as shown in Fig.~\ref{PscHetnet}; by randomly removing macro base stations (MBSs), we simulate a mock emergency situation to study the impact of interference and CRE when the UABSs are deployed.
Subsequently, we explore potential gains in 5th percentile SE (5pSE) from the use of Release-10/11 ICIC techniques for a UABS based PSC network.

The rest of this paper is organized as follows. In Section~\ref{systemModel}, we provide the UABS-based HetNet model, assumptions, and definition of 5pSE as a function of network parameters. The UABSs deployment and ICIC parameter configurations using the genetic algorithm and hexagonal grid UABS model are described in Section~\ref{sec:UabsDeploy}. In Section~\ref{simulation}, we analyze and compare the 5pSE of the HetNet using extensive computer simulations for various ICIC techniques, and finally, the last section provides some concluding remarks.

\section{System Model}
\label{systemModel}
We consider a two-tier HetNet deployment with MBSs and UABSs as shown in Fig.~\ref{PscHetnet}, where all the MBSs and UABS locations are captured in matrices ${\bf X}_{\rm mbs} \in \mathbb{R}^{N_{\rm mbs}\times 3}$  and ${\bf X}_{\rm uabs}\in \mathbb{R}^{N_{\rm uabs}\times 3}$, respectively, where $N_{\rm mbs}$ and $N_{\rm uabs}$ denote the number of MBSs and UABSs within the simulation area, and UABSs are deployed at a fixed height. The MBS and user equipment (UE) locations are each modeled using a two-dimensional Poisson point process (PPP) with intensities $\rm \lambda_{\rm mbs}$ and $\rm \lambda_{\rm ue}$, respectively~\cite{R10, R12}. The UABSs are deployed either at fixed locations in a hexagonal grid, or the locations are optimized using the genetic algorithm. We assume that the MBSs and the UABSs share a common transmission bandwidth, round robin scheduling is used in all downlink transmissions, and full buffer traffic is used in every cell.

The transmit power of the MBS and UABS are $P_{\rm mbs}$ and $P_{\rm uabs}$, respectively, while $K$ and $K^\prime$ are the attenuation factors due to geometrical parameters of antennas for the MBS and the UABS, respectively. Then, the effective transmit power of the MBS is $P^\prime_{\rm mbs} = KP_{\rm mbs}$, while the effective transmit power of the UABS is $P^\prime_{\rm uabs} = K^\prime P_{\rm uabs}$.

An arbitrary UE $n$ is always assumed to connect to the nearest MBS or UABS, where $n\in\{1,2,...,N_{\rm ue}\}$. 
Let the nearest macro-cell of interest (MOI) be at a distance $d_{mn}$ and the nearest UAV-cell of interest (UOI) be at a distance $d_{un}$.  Then, for the $n$th UE the reference symbol received power from the $m$th MOI and the $u$th UOI are given by~\cite{R10}
\begin{align}
     S_{\rm mbs}(d_{mn}) = \frac{P^\prime_{\rm mbs}}{d_{mn}^\delta}, \ S_{\rm uabs}(d_{un}) = \frac{P^\prime_{\rm uabs}}{d_{un}^\delta},
\end{align}
where $\delta$ is the path-loss exponent, and $d_{un}$ depends on the locations of the UABSs that will be dynamically optimized.

\subsection{3GPP Release-10/11 Inter-Cell Interference Coordination}
\label{icicidetails} 
\begin{figure} [t]
\begin{subfigure}[b]{1\linewidth}
\centering
\includegraphics[width=0.85\linewidth]{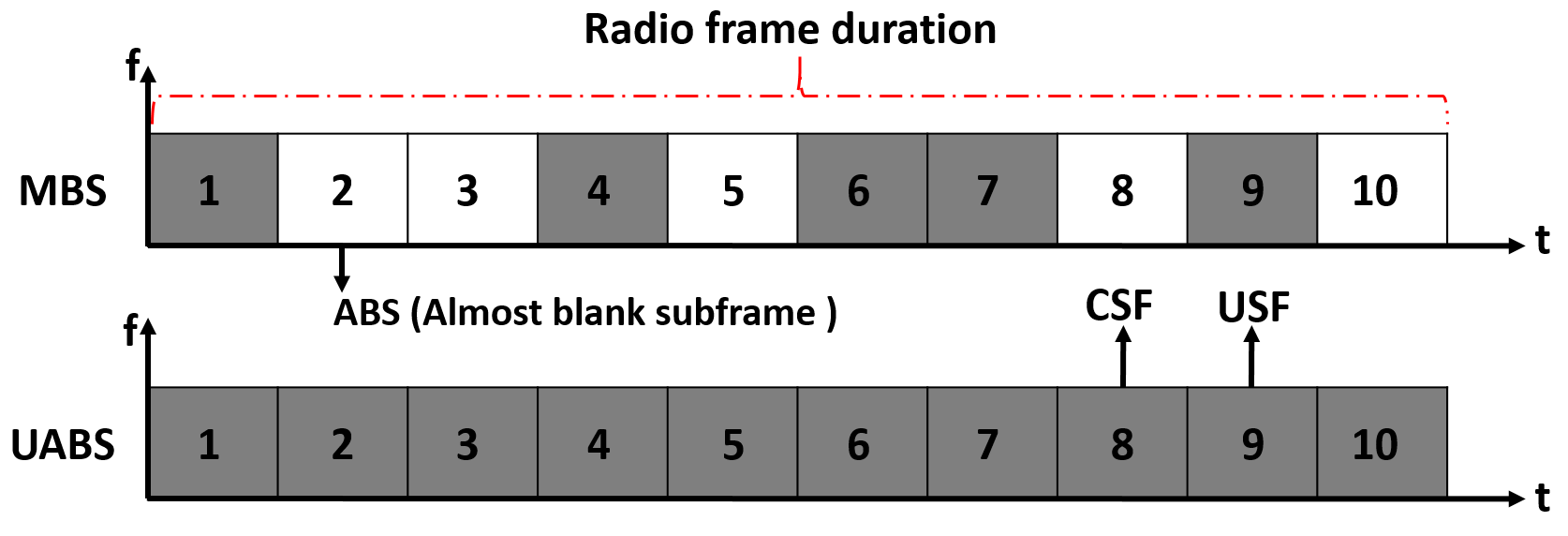}
\caption{3GPP Release-10 eICIC with ABS.}
\label{ReducedPowerFrames}
\end{subfigure}
\begin{subfigure}[b]{1\linewidth}
\centering
\includegraphics[width=0.85\linewidth]{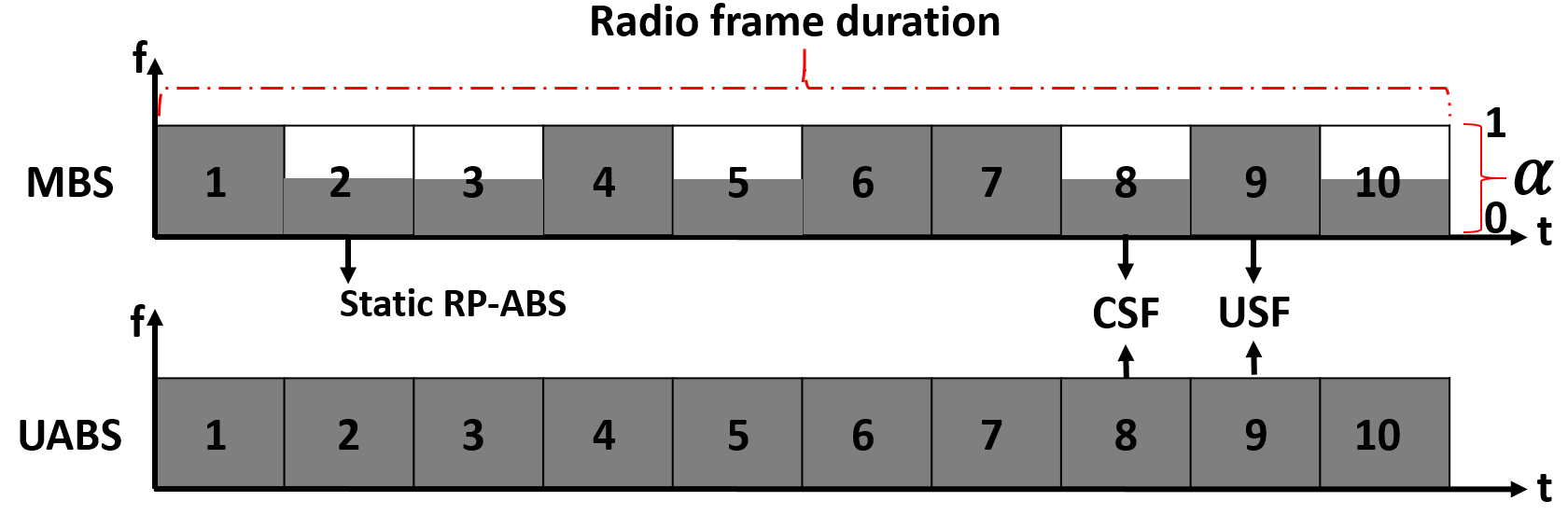}
\caption{3GPP Release-11 FeICIC with reduced power ABS (RP-ABS).}
\label{StaticPrAbs}
\end{subfigure}
\caption{LTE-Advanced frame structures for time-domain ICIC.}\label{Fig2} \vspace{-0.25cm}
\end{figure}

Due to their low transmission power, the UABSs are unable to associate a larger number of UEs compared to that of MBSs. However, by using the cell range expansion (CRE) technique defined in 3GPP Release~8, UABSs can associate a large number of UEs by offloading traffic from MBSs. A negative effect of CRE includes increased interference in the downlink on cell-edge UEs or the UEs in CRE region of the UABS, which is addressed by using ICIC techniques in LTE and LTE-Advanced~\cite{R5, R7, R8}.
3GPP Release-10 introduced a time-domain based enhanced ICIC (eICIC). It uses almost blank subframes (ABS) which require the MBS to completely blank the transmit power on the physical downlink shared channel (PDSCH) resource elements as shown in Fig.~\ref{Fig2}(a). This separates the radio frames into coordinated subframes (CSF) and uncoordinated subframes (USF). 3GPP Release-11 defines further-enhanced ICIC (FeICIC), where the data on PDSCH is still transmitted but at a reduced power level as shown in Fig.~\ref{Fig2}(b). We assume that the ABS and reduced power pattern are shared via the X2 interface, which is a logical interface between the base stations. Implementation of the X2 interface for UABSs is left as a future consideration.

The MBSs can schedule their UEs either in USF or in CSF based on the scheduling threshold $\rho$. Similarly, the UABSs can schedule their UEs either in USF or in CSF based on the scheduling threshold $\rho^\prime$. Let $\beta$ denote the USF duty cycle, defined as the ratio of number of USF subframes to the total number of subframes in a radio frame. Then, the duty cycle of CSFs is $(1-\beta)$. For ease of simulation, the USF duty cycle $\beta$ is fixed at 50\% in this paper for all the MBSs, which is shown in~\cite{R10} to have limited effect on system performance when $\rho$ and $\rho'$ are optimized. Finally, let $0\leq\alpha\leq 1$ denote the power reduction factor in coordinated subframes of the MBS for the FeICIC technique. As two special cases,  $\alpha=0$ corresponds to Release-10 eICIC, while $\alpha=1$ corresponds to no ICIC.

Given the eICIC and FeICIC framework in 3GPP LTE-Advanced as in Fig.~\ref{Fig2}, and following an approach similar to that in~\cite{R10} for a HetNet scenario, the signal-to-interference ratio (SIR) experienced by an arbitrary $n$th UE can be defined for CSFs and USFs for the $m$th MOI and the $u$th UOI as follows:
\begin{align}
\Gamma &= \frac{S_{\rm mbs}(d_{mn})}{S_{\rm uabs}(d_{un}) + Z}\rightarrow {\rm USF\ SIR\ from\ MOI}, \label{Eq:SIR1} \\
\Gamma_{\rm csf} &= \frac{\alpha S_{\rm mbs}(d_{mn})}{S_{\rm uabs}(d_{un}) + Z} \rightarrow {\rm CSF\ SIR\ from\ MOI}, \label{Eq:SIR2} \\
\Gamma^\prime &= \frac{S_{\rm uabs}(d_{un})}{ S_{\rm mbs}(d_{mn}) + Z} \rightarrow {\rm USF\ SIR\ from\ UOI}, \label{Eq:SIR3} \\
\Gamma^\prime_{\rm csf} &= \frac{S_{\rm uabs}(d_{un})}{\alpha S_{\rm mbs}(d_{mn})+ Z} \rightarrow {\rm CSF\ SIR\ from\ UOI},\label{Eq:SIR4}
\end{align}
where $Z$ is the total interference power at a UE during USF or CSF from all the MBSs and UABSs, excluding the MOI and the UOI. In hexagonal grid UABS deployment model (and in~\cite{R10}), locations of the UABSs (and small cells) are fixed. To maximize the 5pSE of the network, we actively consider the SIRs in~\eqref{Eq:SIR1}--\eqref{Eq:SIR4} while optimizing the locations of the UABSs using the genetic algorithm.

\subsection{UE Association and Scheduling}
The cell selection process relies on $\Gamma$ and $\Gamma^\prime$ in \eqref{Eq:SIR1} and \eqref{Eq:SIR3}, respectively, for the MOI and UOI SIRs, as well as the CRE $\tau$. If $\tau\rm\Gamma^\prime$ is less than $\rm \Gamma$, then the UE is associated with the MOI; otherwise, it is associated with the UOI. After cell selection, the MBS-UE (MUE) and UABS-UE (UUE) can be scheduled either in USF or in CSF radio subframes as: 
\begin{align}
& \rm If\ \Gamma\ \textgreater\ \tau\Gamma^\prime\ and\ \Gamma \le\ \rho \rightarrow USF-MUE, \label{Case1}\\
& \rm If\ \Gamma\ \textgreater\ \tau\Gamma^\prime\ and\ \Gamma\ \textgreater\ \rho \rightarrow CSF-MUE, \\
& \rm If\ \Gamma \le \tau\Gamma^\prime\ and\ \Gamma^\prime\ \textgreater\ \rho^\prime \rightarrow USF-UUE,\\
& \rm If\ \Gamma \le \tau\Gamma^\prime\ and\ \Gamma^\prime \le\ \rho^\prime \rightarrow CSF-UUE.\label{Case4}
\end{align}

Once a UE is assigned to an MOI/UOI, and it is scheduled within a USF/CSF, then the SE for this UE can be expressed for the four different scenarios in \eqref{Case1}-\eqref{Case4} as follows:
\begin{align}
C_{\rm usf}^{\rm mbs} &= \frac{\beta {\rm log_2}(1+\Gamma)}{N_{\rm usf}^{\rm mbs}},~\label{Cap1}\\
C_{\rm csf}^{\rm mbs} &= \frac{(1-\beta){\rm log_2}(1+\Gamma_{\rm csf})}{N_{\rm csf}^{\rm mbs}},\\
C_{\rm usf}^{\rm uabs} &= \frac{{\rm log_2}(1+\Gamma^\prime)}{N^{\rm uabs}_{\rm usf}},~\\
C_{\rm csf}^{\rm uabs} &= \frac{ {\rm log_2}(1+\Gamma^\prime_{\rm csf})}{N^{\rm uabs}_{\rm csf}},\label{Cap4}
\end{align}
where $N_{\rm usf}^{\rm mbs}$, $N_{\rm csf}^{\rm mbs}$, $N^{\rm uabs}_{\rm usf}$, and $N^{\rm uabs}_{\rm csf}$ are the number of MUEs and UUEs scheduled in USF and CSF radio subframes, and $\Gamma$, $\Gamma_{\rm csf}$, $\Gamma^\prime$, $\Gamma^\prime_{\rm csf}$ are as in \eqref{Eq:SIR1}-\eqref{Eq:SIR4}. 

In this paper, we consider the use of 5pSE which corresponds to the worst fifth percentile UE capacity among the capacities of all the $N_{\rm ue}$ UEs (calculated based on \eqref{Cap1}-\eqref{Cap4}) within the simulation area. We believe it is a critical metric particularly for PSC scenarios to maintain a minimum QoS level at all the UEs in the environment.
We define the dependency of the 5pSE to UABS locations and ICIC parameters as 
\begin{align}
C_{\rm 5th}\Big({\bf X}_{\rm uabs},{\bf S}_{\rm mbs}^{\rm ICIC},{\bf S}_{\rm uabs}^{\rm ICIC}\Big)~,
\end{align}
where ${\bf X}_{\rm uabs}\in \mathbb{R}^{N_{\rm uabs}\times 3}$ captures the UABS locations as defined earlier, ${\bf S}_{\rm mbs}^{\rm ICIC} = [\boldsymbol{\alpha},\boldsymbol{\rho}]$ $\in \mathbb{R}^{N_{\rm mbs} \times 2}$ is a matrix that captures individual ICIC parameters for each MBS, while ${\bf S}_{\rm uabs}^{\rm ICIC} = [\boldsymbol{\tau},\boldsymbol{\rho'}]$ $\in \mathbb{R}^{N_{\rm uabs} \times 2}$ is a matrix that captures individual ICIC parameters for each UABS. In particular,  
\begin{align}
\boldsymbol{\alpha}=[\alpha_1,...,\alpha_{N_{\rm mbs}}]^T,\quad \boldsymbol{\rho}=[\rho_1,...,\rho_{N_{\rm mbs}}]^T
\end{align}
are $N_{\rm mbs}\times 1$ vectors that include the power reduction factor and MUE scheduling threshold parameters for each MBS. On the other hand, 
\begin{align}
\boldsymbol{\tau}=[\tau_1,...,\tau_{N_{\rm uabs}}]^T, \quad \boldsymbol{\rho}^\prime=[\rho_1^\prime,...,\rho_{N_{\rm uabs}}^\prime]^T
\end{align}
are $N_{\rm uabs}\times 1$ vectors that involve the CRE bias and UUE scheduling threshold at each UABS. 

As noted in Section~\ref{icicidetails}, the duty cycle $\beta$ of ABS and reduced power subframes is assumed to be set to $0.5$ at all MBSs to reduce search space and complexity. 

Considering that the optimum values of the vectors $\boldsymbol{\alpha}$, $\boldsymbol{\rho}$, $\boldsymbol{\rho}'$, and $\boldsymbol{\tau}$ are to be searched over a multi-dimensional space, computational complexity of finding the optimum parameters is prohibitively high. Hence, to reduce system complexity (and simulation runtime) significantly, we consider that the same ICIC parameters are used for all MBSs and all UABSs. In particular, we consider that for $i=1,...,N_{\rm mbs}$ we have~$\alpha_i=\alpha$ and $\rho_i=\rho$, while for $j=1,...,N_{\rm uabs}$ we have $\tau_j=\tau$ and $\rho_j^\prime=\rho^\prime$. Therefore, the dependence of the 5pSE on the UABS locations and ICIC parameters can be simplified as
\begin{align}
C_{\rm 5th}\big({\bf X}_{\rm uabs},\alpha,\rho,\tau,\rho^\prime\big)~,
\end{align}
which we will seek ways to maximize in the next section. We leave the problem of individually optimizing ICIC parameters for the MBSs and UABSs as a future work due to the high computational complexity of the problem.

\section{UABS Deployment Optimization}
\label{sec:UabsDeploy}
We consider that the cell-edge user SE is captured by the 5pSE of the cumulative distribution function of the user throughput, which we will use as a metric to measure the overall network performance. In this section, we discuss the UABS deployment using the genetic algorithm (GA) and the hexagonal grid model, where we use the 5pSE as an optimization metric to maximize for both scenarios.

\begin{figure} [t]
\centering
\includegraphics[width=0.85\linewidth]{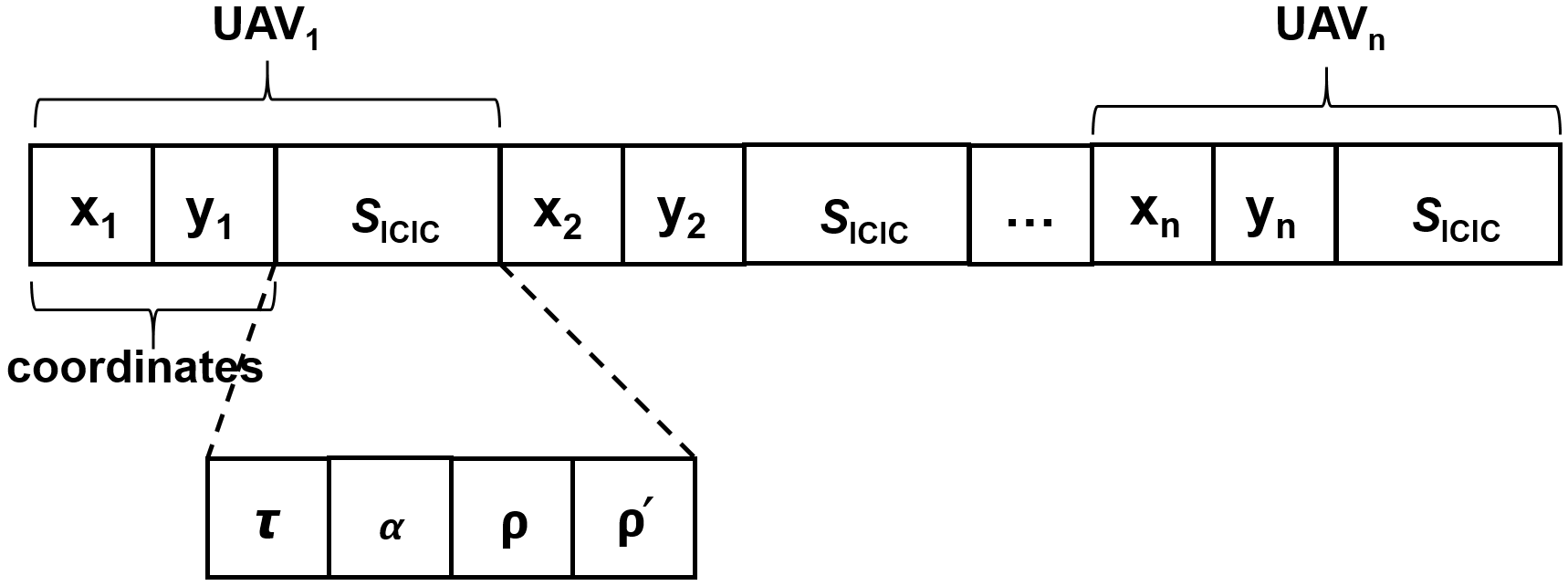}
\caption{An example of a chromosome for FeICIC simulation, where the UABS locations, ICIC parameter $\tau$, $\alpha$, $\rho$, and $\rho^\prime$ are optimized. The ICIC parameter $\beta$ is not optimized and is fixed at 50\% duty cycle.}
\label{GAChromosome} \vspace{-0.2cm}
\end{figure}

\subsection{Genetic Algorithm based UABS Deployment Optimization}
\label{GaSection}
The GA is a population-based optimization technique that can search a large environment simultaneously to reach an optimal solution \cite{merwaday2016improved}. In this paper, the UABS coordinates and the ICIC parameters constitute the GA population, and a subsequent chromosome is illustrated in Fig. \ref{GAChromosome}. Listing \ref{GaListing} describes the main steps used to optimize the UABS locations and ICIC parameters while computing the 5pSE.

We apply the GA to simultaneously optimize the UABS locations and ICIC parameters to maximize the 5pSE of the network over a given geographical area of interest. The location of each UABS within a rectangular simulation area is given by $(x_i,y_i)$ where $i \in \{ 1,2,...,N_{\rm uabs}\}$. The UABS locations and the ICIC parameters that maximize the 5pSE objective function can be calculated as
\begin{align}
\big[\hat{\bf X}_{\rm uabs},\hat{\alpha},\hat{\rho},&\hat{\tau},\hat{\rho^\prime}\big]= \nonumber\\
&\arg\underset{{\bf X}_{\rm uabs},\alpha,\rho,\tau,\rho^\prime}{\max} C_{\rm 5th}\big({\bf X}_{\rm uabs},\alpha,\rho,\tau,\rho^\prime\big). \label{GA_Optim}
\end{align}
Since searching for optimal ${\bf X}_{\rm uabs}$ and ICIC parameters through a brute force approach is computationally intensive, in this paper, we use the GA to find optimum UABS locations and the best-fit ICIC parameters $\tau$, $\alpha$, $\rho$, and $\rho^\prime$.

\begin{lstlisting}[caption=Steps for optimizing population using GA., basicstyle=\scriptsize, language=R, breaklines=true, numbers=none, frame=single, showstringspaces=false, xleftmargin=0.2cm, linewidth=8.7cm,label=GaListing]
Input:
Population: set of UABS locations and ICIC parameters
FITNESS function: 5pSE of the network
Output:
Args: Best individuals of ICIC parameters and
      highest 5pSE
Method:
NewPopulation <- empty set
StopCondition: Number of iterations = 6
SELECTION: Roulette wheel selection method
while(! StopCondition)
{
  for i = 1 to Size do
  {
    Parent1 <- SELECTION(NewPopulation,FITNESS function)
    Parent2 <- SELECTION(NewPopulation,FITNESS function)
    Child <- Reproduce(Parent1, Parent2)
    if(small random probability)
    {
     child <- MUTATE(Child)
     add child to NewPopulation
    }
  }
  EVALUATE(NewPopulation, FITNESS function);
  Args <- GetBestSolution(NewPopulation)
  Population <- Replace(Population, NewPopulation)
}
\end{lstlisting}

\begin{table}[t]
\caption{Simulation parameters.}
\label{tab:SysParams}
\centering
\footnotesize
\begin{tabular}{|p{4.85cm}|p{3.25cm}|}
\hline
{\bf Parameter} & {\bf Value}  \\ \hline
MBS and UE intensity & $4$ per km$^2$ and $100$ per km$^2$ \\ \hline
MBS and UABS transmit powers & $46 {\rm\ dBm}$ and $30 {\rm\ dBm}$\\ \hline
Path-loss exponent & $4$\\ \hline
Altitude of UABSs & $121.92 {\rm\ m}$ ($400 {\rm\ feet}$)\\ \hline
Simulation area & $10 \times 10 {\rm\ km^2}$\\ \hline
GA population size and generation number & $60$ and $100$\\\hline
GA crossover and mutation probabilities & $0.7$ and $0.1$\\\hline
Cell range expansion ($\tau$) in dB & $0$ to $15$ dB \\\hline
Power reduction factor for MBS during ($\alpha$)  &  $0$ to $1$  \\ \hline
Duty cycle for the transmission of USF ($\beta$)  &  $0.5$ or 50\%   \\ \hline
Scheduling threshold for MUEs ($\rho$)            &  $20$ dB to $40$ dB \\ \hline
Scheduling threshold for UUEs ($\rho\prime$)      &  $-20$ dB to $-10$ dB \\ \hline
MBS destroyed sequence & $50\%$ and $97.5\%$ \\\hline
PSC LTE Band~14 center frequency & 763 MHz for downlink and 793 MHz for uplink \\ \hline
\end{tabular}
\vspace{-2mm}
\end{table}

\subsection{UABS Deployment on a Hexagonal Grid}
\label{HexSection}
As a lower complexity alternative to optimizing UABS locations, we consider deploying the UABSs on a hexagonal grid, where the position of the UABSs are deterministic. We assume that the UABSs are placed within the rectangular simulation area regardless of the existing MBS locations. The 5pSE for this network is determined by using a brute force technique as described in the pseudo-code~\ref{HexListing} which only considers optimization of the ICIC parameters captured through the matrix ${\bf S}_{\rm ICIC}$. The optimized ICIC parameters that maximize the 5pSE can then be calculated as:
\begin{equation}
\big[\hat{\alpha},\hat{\rho},\hat{\tau},\hat{\rho^\prime}\big]=\arg\underset{\alpha,\rho,\tau,\rho^\prime}{\max} ~ C_{\rm 5th}\big({\bf X}^{\rm (hex)}_{\rm uabs},\alpha,\rho,\tau,\rho^\prime\big),\label{Hex_Optim}
\end{equation}
where ${{{\bf X}_{\rm uabs}^{\rm (hex)}}}$ are the fixed and known hexagonal locations of the deployed UABSs within the simulation area.

\begin{lstlisting}[caption=Steps for computing 5pSE for hexagonal grid deployment., basicstyle=\scriptsize, language=R, breaklines=true, numbers=none, frame=single, showstringspaces=false, xleftmargin=0.2cm, linewidth=8.7cm, label=HexListing]
Input: set of UABS locations and ICIC parameters
Output: SE: 5pSE for the network
Method:
StopCondition: Number of iterations = 100
while(! StopCondition)
{
  Generate UABS locations
  for t = 1 to ICICParms.tau[t] do
  {
    for a = 1 to ICICParms.alpha[a] do
    {
      for r = 1 to ICICParms.rho[r] do
      {
        for p = 1 to ICICParms.rhoprime[p] do
        {
          SE = Calc5thPercentileSE(nodal locations, nodal Tx powers, path-loss, tau, beta, alpha, rho, rhoprime)
        }
      }
    }
  }
}
\end{lstlisting}

\begin{figure*} [t]
\vspace{-2mm}
\centering
\begin{subfigure}[b]{0.3\textwidth}
\label{HexPppNIM1}
\includegraphics[width=1\textwidth]{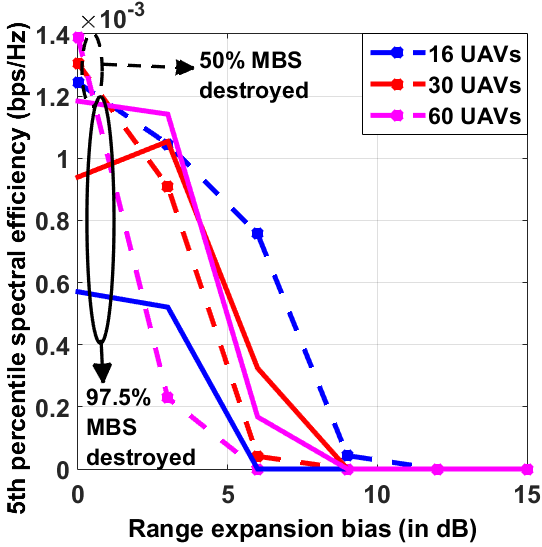}
\caption{5pSE without any ICIC.}
\end{subfigure}
\begin{subfigure}[b]{0.3\textwidth}
\label{HexPppeICIC}
\includegraphics[width=1.01\textwidth]{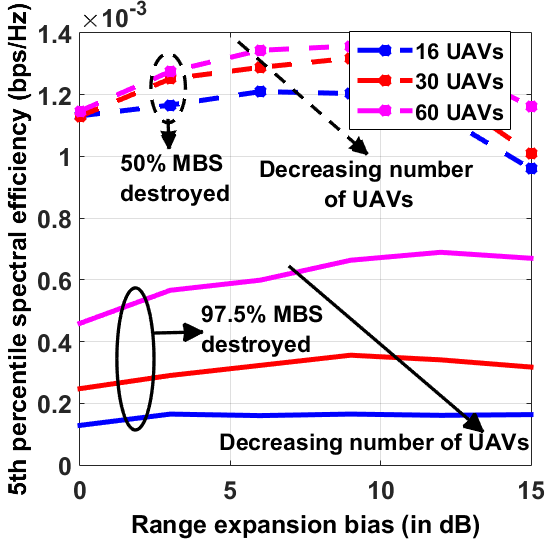}
\caption{5pSE with eICIC.}
\end{subfigure}
\begin{subfigure}[b]{0.3\textwidth}
\label{HexPppFeICIC}
\includegraphics[width=1\textwidth]{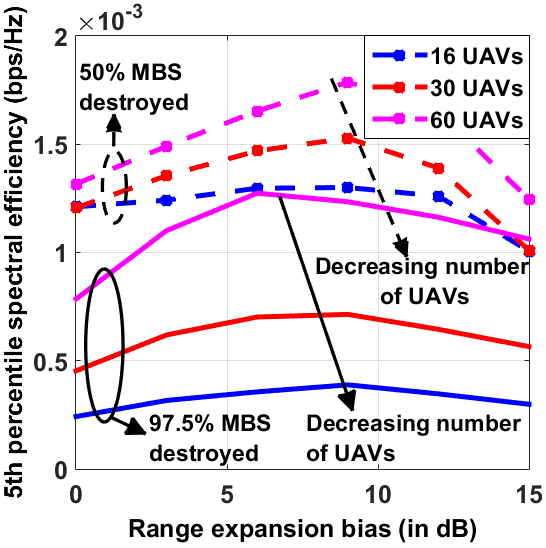}
\caption{5pSE with FeICIC.}
\end{subfigure}
\caption{5pSE versus CRE for eICIC and FeICIC techniques, when the UABSs are deployed on a hexagonal grid.}
\label{fig:SE_PPPUE_HexDep}
\end{figure*}

\section{Simulation Results}
\label{simulation}
In this section, using Matlab based computer simulations, we compare the 5pSE with and without ICIC techniques while considering different UABS deployment strategies. Unless otherwise specified, the system parameters for the simulations are set to the values in Table~\ref{tab:SysParams}.

\subsection{5pSE with UABSs Deployed on a Hexagonal Grid}
The variations in 5pSE with respect to CRE, when the UABSs are deployed on a hexagonal grid and utilizing optimized ICIC parameters (see~\eqref{Hex_Optim} and Listing~\ref{HexListing}) are shown in Fig.~\ref{fig:SE_PPPUE_HexDep}. With no CRE, the number of UEs associated with the UABSs and the interference experienced by these UEs is minimal. With the no-ICIC mechanism (NIM) the peak value for the 5pSE is observed at around 0~dB CRE as seen in Fig.~\ref{fig:SE_PPPUE_HexDep}(a). On the other hand, the 5pSE for ICIC techniques at 0 dB CRE are relatively lower as seen in Fig.~\ref{fig:SE_PPPUE_HexDep}(b) and Fig.~\ref{fig:SE_PPPUE_HexDep}(c), due to blank subframes at the MBSs for eICIC, and power reduction of the CSFs at the MBSs for FeICIC.

\begin{figure} [t]
\centering
\includegraphics[width=0.65\linewidth]{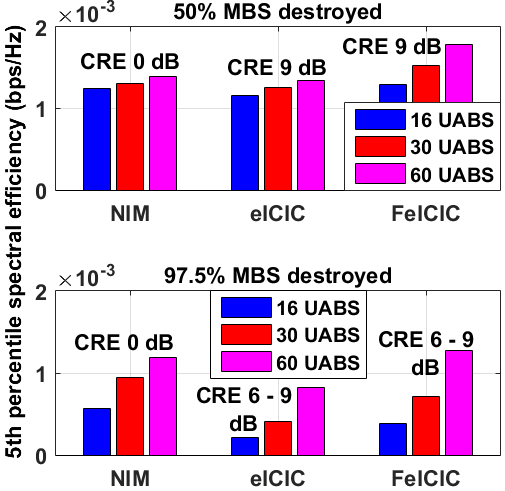}
\caption{Peak observations for the 5pSE, when the UABS are deployed on a fixed hexagonal grid.}
\label{fig:HexGridAna}
\vspace{-3mm}
\end{figure}

As the CRE increases, the number of UEs associated with the UABSs increases and so does the interference experienced by these UEs. Hence, with NIM the 5pSE decreases with increasing CRE as seen in Fig.~\ref{fig:SE_PPPUE_HexDep}(a). On the other hand, the ICIC techniques observe improvement in SE performance. The peak values of the 5pSE for the ICIC techniques is observed when the CRE is between $6-9$~dB. This influence of CRE on the 5pSE for NIM and ICIC is summarized in Fig.~\ref{fig:HexGridAna}.

Overall, the 5pSE for the network is higher when larger numbers of UABSs are deployed and when fewer MBSs are destroyed. Also, the 5pSE decreases with the increasing number of destroyed MBSs as seen in Fig.~\ref{fig:SE_PPPUE_HexDep}.

\subsection{5pSE with GA Based UABS Deployment Optimization}
\begin{figure}
\centering
\vspace{-3mm}
\begin{subfigure}[b]{0.24\textwidth}
\label{HexPppeICIC}
\includegraphics[width=0.97\textwidth]{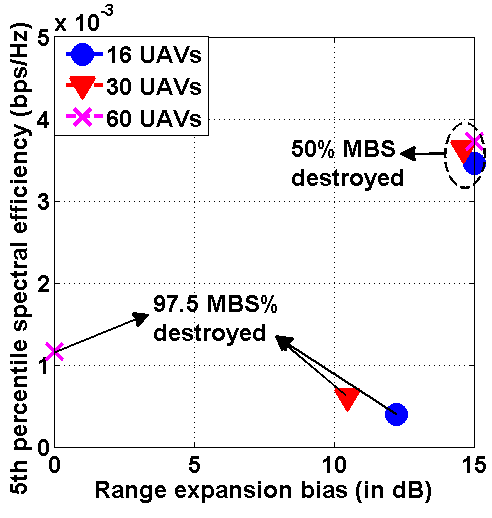}
\caption{5pSE with eICIC.}
\end{subfigure}
\begin{subfigure}[b]{0.24\textwidth}
\label{HexPppFeICIC}
\includegraphics[width=1\textwidth]{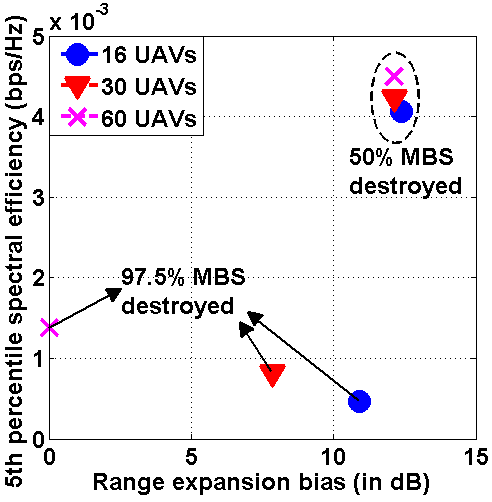}
\caption{5pSE with FeICIC.}
\end{subfigure}
\caption{Peak 5pSE versus optimized CRE for eICIC and FeICIC techniques, when the UABS locations and ICIC parameters are optimized using the GA.}
\label{fig:SE_PPPUE_GaDep} 
\vspace{-1mm}
\end{figure}

Using the UABS locations and ICIC parameters optimized through the GA as in~\eqref{GA_Optim} and Listing~\ref{GaListing}, we plot the peak 5pSE for the network with respect to the optimized CRE value in Fig.~\ref{fig:SE_PPPUE_GaDep}. In the GA based simulations, the optimum CRE value is directly related to the locations of the UABSs with respect to the MBSs, the number of UEs offloaded to the UABSs, and the amount of interference observed by the UEs.

Consider first that the $50\%$ of the MBSs are destroyed, which implies that there are still a large number of MBSs present and the interference from these MBSs is substantial. Hence, offloading a large number of UEs from MBSs to UABSs with higher values of CRE and ICIC is necessary for achieving better 5pSE gains as shown in Fig.~\ref{fig:SE_PPPUE_GaDep}(a) and Fig.~\ref{fig:SE_PPPUE_GaDep}(b) for eICIC and FeICIC, respectively.

When most of the infrastructure is destroyed (i.e., when $97.5\%$ of the MBSs destroyed), the interference observed from the MBSs is limited and larger number of UEs need to be served by the UABSs. Therefore, with fewer UABSs deployed, higher CRE are required to serve a larger number of UEs and achieve better 5pSE. On the other hand, when a larger number of UABSs are deployed, smaller values of CRE will result in better 5pSE gains. We record these behavior in Fig.~\ref{fig:SE_PPPUE_GaDep}(a) and Fig.~\ref{fig:SE_PPPUE_GaDep}(b) for eICIC and FeICIC, respectively.

\begin{figure}
\centering
\vspace{-5mm}
\begin{subfigure}[b]{0.35\textwidth}
\label{HexPppeICIC}
\includegraphics[width=0.85\textwidth]{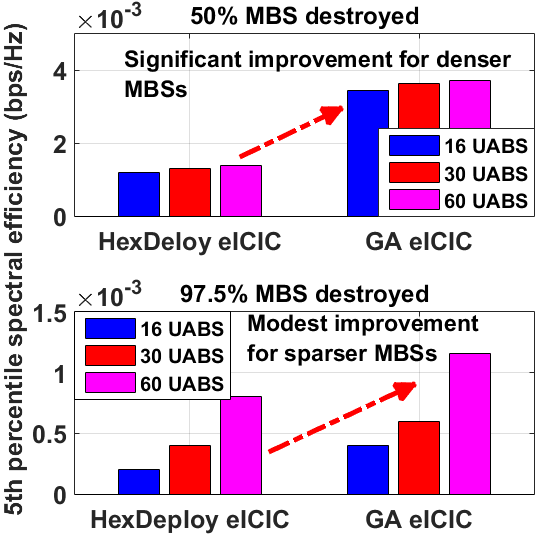}
\caption{5pSE with eICIC.}
\end{subfigure}
\begin{subfigure}[b]{0.35\textwidth}
\label{HexPppFeICIC}
\includegraphics[width=0.85\textwidth]{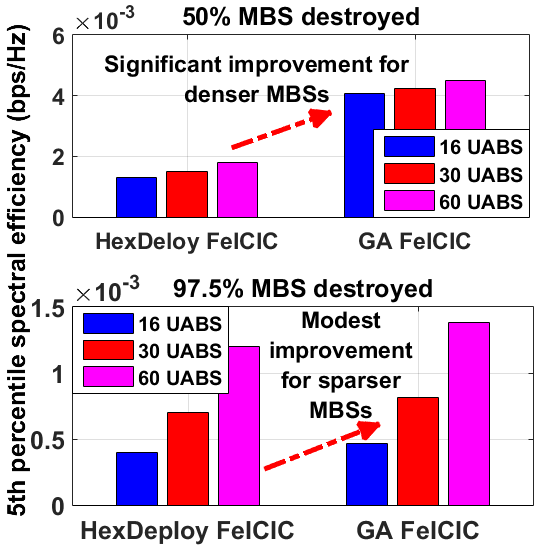}
\caption{5pSE with FeICIC.}
\end{subfigure}
\caption{5pSE comparisons for eICIC and FeICIC techniques, when the UABS locations are optimized using the GA and when the UABSs are deployed in a fixed hexagonal grid.}
\label{fig:SE_HexVsGa}
\vspace{-3mm}
\end{figure}

\subsection{Performance Comparison Between Fixed (Hexagonal) and Optimized UABS Deployment with eICIC and FeICIC}

We summarize our key results from earlier simulations in Fig.~\ref{fig:SE_HexVsGa} to compare the key trade-offs between fixed (hexagonal) deployment and GA based deployment of UABSs. The comparison of Fig.~\ref{fig:SE_PPPUE_HexDep} and Fig.~\ref{fig:SE_PPPUE_GaDep} show that the optimized deployment of UABSs provides a better 5pSE than the UABSs deployed on a fixed hexagonal grid, which are also reflected in the comparative analysis in Fig.~\ref{fig:SE_HexVsGa} which considers an optimized CRE. Moreover, Fig.~\ref{fig:SE_HexVsGa} shows that the 5pSE gains from the optimization of UABS locations are more significant when $50\%$ of the MBSs are destroyed and less significant when $97.5\%$ of the MBSs are destroyed. 

When $50\%$ MBSs are destroyed, there are still a large number of MBSs present which causes substantial interference. Hence, in such interference driven scenario it is important to optimize the locations of the UABSs, and use of larger number of UABSs provide only marginal gains in the 5pSE.

On the other hand, with $97.5\%$ of the MBSs destroyed, the interference from the MBSs is small, and deploying the UABSs on a hexagonal grid will perform close to optimum UABS deployment. The difference between the hexagonal deployment and optimized deployment is especially small for the FeICIC scenario where power reduction factor $\alpha$ in the MBS CSFs provides an additional optimization dimension for improving the 5pSE. Use of a larger number of UABSs when $97.5\%$ of the MBSs are destroyed is also shown to provide significant gains in the 5pSE, in contrast to modest gains in the 5pSE when $50\%$ of the MBSs are destroyed.

\section{Concluding remarks}
\label{conclusion}

In this article, we show that the mission-critical communications could be maintained by deploying UABSs in the event of any damage to the public safety infrastructure. Through simulations, we compare and analyze the 5pSE of the network when the UABSs are deployed on a hexagonal grid and when placed optimally using the GA. Our analysis shows that the deployment of the UABSs on a hexagonal grid is close to optimal when the observed interference is limited. In the presence of substantial interference, the GA approach is more effective for deploying UABSs. Finally, we observe that the HetNets, with reduced power subframes (FeICIC) yields better 5pSE than that with almost blank subframes (eICIC). Future research directions include optimizing the UABS locations using learning techniques and developing a path-planning algorithm for UABS placement.

\section*{Acknowledgment}
This research was supported in part by NSF under the grants AST-1443999 and CNS-1453678. The authors would like to thank A. Merwaday for his helpful feedback.

\ifCLASSOPTIONcaptionsoff
  \newpage
\fi



%
\bibliographystyle{IEEEtran}
\bibliography{Citations}

\end{document}